\documentclass{aa}  
\usepackage{graphicx}
\usepackage[varg]{txfonts}
\usepackage{natbib}
\usepackage{gensymb}
\usepackage{multirow}
\usepackage{color}
\usepackage[normalem]{ulem}
\usepackage[]{siunitx}
\usepackage{rotating}

\newcommand\ions[2]{{#1}\,{\sc #2}} 

\begin{document}

\title{Two stellar populations with different metallicities in the low-mass globular cluster Gran 5}

\author{Dongwook Lim\inst{1}
  \and Sang-Hyun Chun\inst{2}
  \and Young-Wook Lee\inst{1}
  \and Chul Chung \inst{1}
  \and Andreas J. Koch-Hansen\inst{3}
  \and Seungsoo Hong\inst{4} 
  }

\authorrunning{D. Lim et al.}
\titlerunning{}


\institute{
    Center for Galaxy Evolution Research \& Department of Astronomy, Yonsei University, 50 Yonsei-ro, Seoul 03722, Republic of Korea,
    \email{dwlim@yonsei.ac.kr}
    \and Korea Astronomy and Space Science Institute, 776 Daedeokdae-ro, Yuseong-gu, Daejeon 34055, Republic of Korea
    \and Zentrum f\"ur Astronomie der Universit\"at Heidelberg, Astronomisches Rechen-Institut, M\"onchhofstr. 12-14, 69120 Heidelberg, Germany
    \and Department of Physics and Astronomy, Seoul National University, 1 Gwanak-ro, Gwanak-gu, Seoul 08826, Republic of Korea
    }

\date{Received / Accepted }

\abstract 
{With the increasing number of discoveries of globular clusters in the inner Milky Way, the need for spectroscopic confirmation and further investigation of their stellar populations and chemodynamical properties has become crucial.} 
{Gran~5 is a newly reported low-mass globular cluster located close to the Galactic center, and it is thought to be an accreted object associated with the $Gaia$-Enceladus structure. 
This study aims to investigate the stellar populations of Gran~5 and their detailed chemical properties.}
{We performed high-resolution near-infrared spectroscopy on seven stars in the field of Gran~5 using IGRINS on the Gemini-South telescope.} 
{We identified six stars as cluster members and reveal that they are divided into two stellar populations with different metallicities, with mean [Fe/H] values of $-0.76$~dex and $-0.55$~dex, respectively. 
In addition, the chemodynamical properties of Gran~5 agree with those of in situ globular clusters.}
{Our findings represent the first detection of two stellar populations with different metallicities in a low-mass globular cluster. 
This suggests that the metallicity variation in Gran~5 may have arisen from processes different from those in other globular clusters with metallicity variation, or that it may have lost a substantial amount of its initial mass during its evolution.}
\maketitle


\section{Introduction} \label{sec:intro}
Globular clusters (GCs) are among the most widely studied objects in astronomy due to their luminosity and stellar fossil records.
About three decades ago, the research field related to Milky Way GCs experienced a paradigm shift through the discovery that each cluster hosts multiple stellar populations \citep[e.g.,][]{Lee1999, Bedin2004}. 
The presence of multiple populations was supported by various forms of evidence obtained through diverse photometric and spectroscopic observations, including the splitting of the main sequence and red giant branch (RGB), as well as the Na-O anti-correlation (see \citealt{Bastian2018, Gratton2019, Cassisi2020, Milone2022} and reference therein).  
Ongoing in-depth studies continue to examine both the specific and general properties of GCs using a range of techniques, including Hubble Space Telescope photometry and high-resolution spectroscopy \citep[e.g.,][]{D'Antona2022, CabreraGarcia2024}. 
This paradigm shift not only influences our insights into GC evolution, but also enhances our understanding of their contribution to the formation of the Milky Way because certain chemical anomalies observed in field stars can only be formed in the environment of GCs that harbor multiple stellar populations \citep[see, e.g.,][]{Koch2019, Lim2021b}. 
Research on GCs and their connection to the Milky Way is now actively pursued, particularly based on large photometric and spectroscopic surveys such as the Vista Variables in the Via Lactea \citep[VVV;][]{Minniti2010} and the Apache Point Observatory Galactic Evolution Experiment \citep[APOGEE;][]{Majewski2017}.

As diverse survey observations for the Milky Way stellar component are conducted, our general understanding of GCs is also evolving. 
One major achievement is the discovery of numerous new GC candidates in the Galactic plane, primarily through near-infrared (NIR) photometric surveys such as VVV and the Wide-field Infrared Survey Explorer \citep[WISE;][]{Wright2010}, accompanied by precise astrometric solutions and proper motion data provided by the $Gaia$ mission \citep{GaiaCollaboration2016}. 
Over the past five years, more than one hundred new candidates have been reported \citep[e.g.,][]{Borissova2018, Camargo2018, Minniti2021, Garro2022}. This marks a significant increase compared to the number of previously known GCs in the Milky Way (e.g., 157 GCs are listed in \citealt{Harris2010}). 
Because GCs are expected to play an important role in the formation of the Milky Way, it is crucial for advancing our understanding of galaxy formation and evolution to study the numerous new GCs and their detailed properties. 
In this respect, while comprehensive studies of newly discovered GCs have been conducted by \citet{Bica2024} and \citet{Garro2024}, who each compiled data for 41 and 37 new GCs, detailed investigations of individual GCs are also ongoing \citep[e.g.,][]{Kunder2021, Romero-Colmenares2021, Pace2023}.

However, before we conduct detailed research on the newly discovered GC candidates, it is essential to confirm whether they are indeed stellar clusters. 
For instance, \citet{Lim2022} performed spectroscopic observations of stars in Camargo~1103 and Camargo~1106, determining that they did not match the expected characteristics of GCs as previously predicted from photometric observations \citep{Camargo2018}.
The most direct method to ascertain the status of these candidates as genuine GCs involves high-resolution spectroscopy of individual stars, which verifies the similarity in line-of-sight velocity and chemical composition of member stars that are selected based on their positions and proper motions.
In the case of stars toward the Galactic bulge, where foreground and background extinction are severe, NIR spectroscopy offers a significant advantage.

In this study, we perform high-resolution NIR spectroscopy on seven stars in the field of Gran~5, using the Immersion Grating Infrared Spectrometer \citep[IGRINS;][]{Mace2018} to investigate its stellar populations and chemodynamical properties. 
Gran~5 is one of the new GCs discovered by \citet{Gran2022}, who reported five new GC candidates based on $Gaia$ and VVV data.  
They measured the radial velocity (RV) and metallicity from Ca triplet lines around 8,500~$\AA$ obtained with the Multi Unit Spectroscopic Explorer (MUSE) instrument, resulting in mean values of $-90.40$~km~s$^{-1}$ for the RV and $-1.56$~dex for [Fe/H]. 
These estimates were recently revised to $-59.19$~km~s$^{-1}$ and $-1.02$~dex by \citet{Gran2024} through a full spectral synthesis using the same MUSE data. 
While \citet{Gran2022} suggested that Gran~5 is an ``accreted'' bulge GC that appears to be related to the $Gaia$-Enceladus structure \citep{Helmi2018}, \citet{Belokurov2024} classified it as an ``in situ'' GC. 
A more detailed chemodynamical approach is therefore required for a better understanding of the origin of Gran~5.
In addition, the presence or absence of multiple stellar populations in this GC is one of the intriguing aspects, as Gran~5 is expected to be a low-mass GC \citep[$\sim10^4$ M$_{\odot}$;][]{Gran2022} that is located very close to the Galactic center.

The structure of this paper is as follows. 
The observational procedures and data reduction methods are provided in Sect.~\ref{sec:obs_reduction}. Section~\ref{sec:member_spec} covers the determination of cluster membership and the measurement of atmospheric parameters and chemical abundances for the identified member stars. 
Section~\ref{sec:result_chem} presents the detailed chemical properties of the observed stars, and Sect.~\ref{sec:orbit} describes the overall chemodynamical properties of Gran~5.
Finally, we discuss the implications of our findings in Sect.~\ref{sec:discussion}.


\section{Observation and data reduction} \label{sec:obs_reduction}
\subsection{Observation} \label{sec:sub:obs}
\begin{figure}
\centering
   \includegraphics[width=0.45\textwidth]{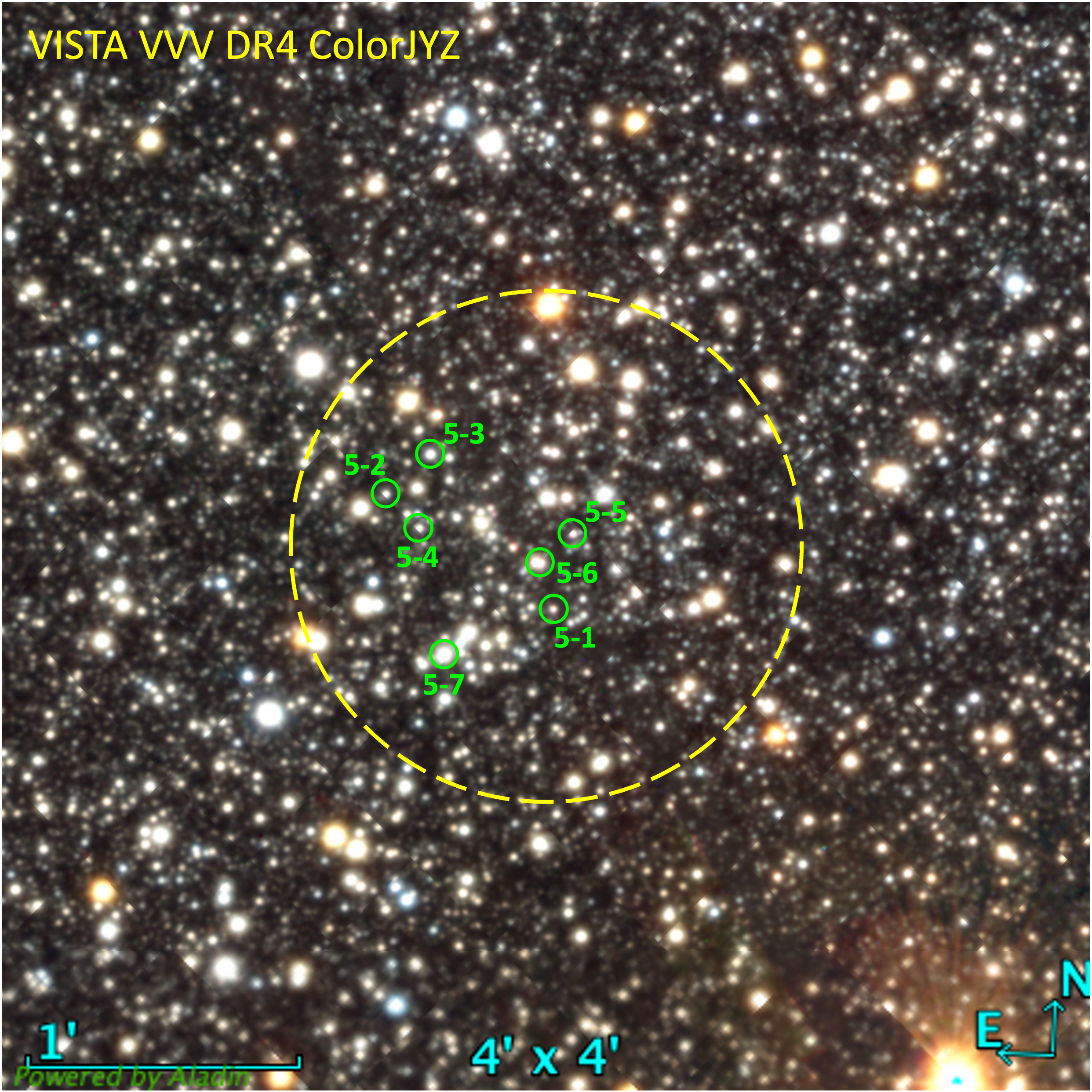}
     \caption{$4\arcmin \times 4\arcmin$ field on the VVV DR4 image of Gran~5.
     Our seven spectroscopic target stars are marked by green circles. 
     The yellow circle represents the half-light radius (0.94$\arcmin$) of Gran~5, as estimated by \citet{Gran2022}.}
     \label{fig:image}
\end{figure}

\begin{figure*}
\centering
   \includegraphics[width=1.0\textwidth]{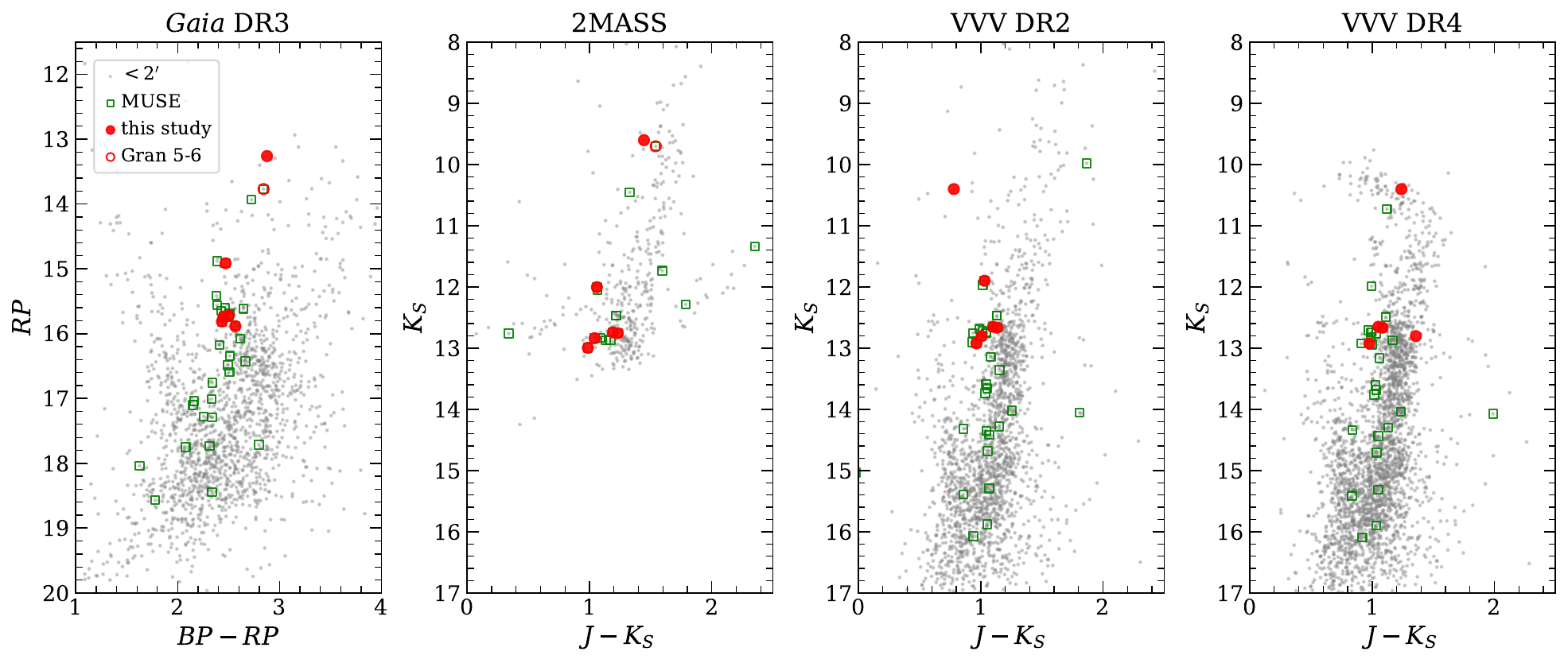}
     \caption{Our target stars are marked with red circles in CMDs that are cross-matched with the $Gaia$ DR3, 2MASS, VVV DR2, and VVV DR4 catalogs.
     The open red circle represents Gran~5-6, which is identified as a nonmember star in this study. 
     The green squares indicate member candidates identified using MUSE data by \citet{Gran2024}, and gray circles are background stars located within 2$\arcmin$ from the center of Gran~5.
     Gran~5-3 and Gran~5-6 are not identified in the VVV DR4 CMD due to the absence of $J$ magnitude, and Gran~5-6 is not marked either in the VVV DR2 CMD.
     In addition, the $J - K_{s}$ color of Gran~5-7 has shifted to the redder side in VVV DR4 compared to VVV DR2.
     }
     \label{fig:cmd}
\end{figure*}

The reported location of Gran~5 is very close to the Galactic center ($l$ = 4.459$\degree$; $b$ = 1.838$\degree$), which implies that this GC candidate is embedded among numerous Galactic field stars and in the interstellar medium. 
This challenging environment surrounding Gran~5 makes the selection of member stars for high-resolution spectroscopy both difficult and crucial.
\citet{Gran2022}, who first discovered Gran~5, identified 76 potential member stars based on coordinates, color, and proper motion data from $Gaia$ and VVV. 
Subsequently, more reliable member candidates were selected based on RV criteria, with a deviation of less than 5~km~s$^{-1}$ from the mean, using follow-up spectroscopy with MUSE at the Very Large Telescope (VLT).
Although these observations do not guarantee the identification of cluster members because the spectral resolution of MUSE is limited ($R\sim3,000$), we selected our targets for high-resolution spectroscopy from among those identified as member candidates in \citet{Gran2022} to increase the likelihood of an accurate member selection.
Finally, seven bright giant stars were selected, considering their locations in the color-magnitude diagrams (CMDs) using data from $Gaia$ DR3 \citep{GaiaCollaboration2023}, from the Two Micron All Sky Survey \citep[2MASS;][]{Skrutskie2006}, and from VVV DR2 and DR4 \citep{Saito2012}. 
These targets are marked in the VVV DR4 color image in Fig.~\ref{fig:image} and in the ($RP,~BP - RP$) and ($K_{s},~J - K_{s}$) CMDs in Fig.~\ref{fig:cmd}. 
We list the IDs for the observed stars, ranging from Gran~5-1 to Gran~5-7, and their coordinates, proper motions, and magnitudes, updated from the recent $Gaia$ DR3, in Table~\ref{tab:target}. 
No RV information is provided for our target stars in $Gaia$ DR3.
We also provide the cross-matched IDs of the sample stars with those from \citet{Gran2024}, who reanalyzed MUSE data from \citet{Gran2022}. 
However, one of our sample stars, Gran~5-7, is not included in the list from \citet{Gran2024}.

\begin{table*}
\small
\setlength{\tabcolsep}{3pt}
\caption{Target information}
\label{tab:target} 
\centering                                    
\begin{tabular}{ccccccccccc}   
\hline\hline        
\multirow{2}{*}{ID} & \multirow{2}{*}{$Gaia$ DR3 source ID}   & $\alpha$ (J2000)  & $\delta$ (J2000)  & $\mu_{\alpha}$    & $\mu_{\delta}$    & $Gaia$~${BP}$   & $Gaia$~${RP}$ & 2MASS~${Ks}$  & VVV~${Ks}$    & \multirow{2}{*}{Cross-ID}        \\
                    &                                         & [deg]             & [deg]             & [mas yr$^{-1}$]   & [mas yr$^{-1}$]   & [mag]           & [mag]         & [mag]         & [mag]         &                                  \\
\hline
Gran~5-1 & 4068467527032867328 & 267.227476 & $-$24.173905 &  $-$5.467 &  $-$9.334 &  18.45 &  15.88 &  12.76 &  12.69 & 042 \\
Gran~5-2 & 4068469004597425920 & 267.238732 & $-$24.166819 &  $-$5.499 &  $-$9.342 &  18.25 &  15.81 &  13.00 &  12.95 & 137 \\
Gran~5-3 & 4068469000193042944 & 267.235781 & $-$24.164418 &  $-$5.420 &  $-$9.347 &  17.38 &  14.91 &  12.01 &  11.92 & 156 \\
Gran~5-4 & 4068469004493522688 & 267.236450 & $-$24.168945 &  $-$5.511 &  $-$9.316 &  18.22 &  15.72 &  12.84 &  12.83 & 106 \\
Gran~5-5 & 4068467527025031552 & 267.226197 & $-$24.169361 &  $-$5.586 &  $-$9.625 &  18.19 &  15.74 &  12.75 &  12.68 & 100 \\
Gran~5-6 & 4068467522724283392 & 267.228410 & $-$24.171135 &  $-$4.976 &  $-$9.321 &  16.61 &  13.77 &   9.70 &   8.68 & 078 \\
Gran~5-7 & 4068468248683158016 & 267.234857 & $-$24.176719 &  $-$5.350 &  $-$9.350 &  16.14 &  13.26 &   9.60 &   9.36 & -- \\
\hline                                             
\end{tabular}
\tablefoot{Column~10 refers to VVV DR4. Column~11 lists the ID in \citet{Gran2024}, cross-matched with the sample stars.}
\end{table*}

High-resolution NIR spectroscopic data for the seven selected stars were obtained using the IGRINS instrument at the Gemini-South telescope. 
The data cover the H-band (1.45 $-$ 1.8 $\mu$m) and K-band (1.95 $-$ 2.45 $\mu$m) wavelength ranges with a spectral resolving power of $R \sim 45,000$ \citep{Park2014, Mace2018}. 
Five of the stars were observed during the 2023A semester under programs GS-2023A-Q-126 (Gran~5-1, Gran~5-2, and Gran~5-3) and GS-2023A-Q-216 (Gran~5-4 and Gran~5-5), and the remaining two bright stars (Gran~5-6 and Gran~5-7) were observed under the Band 3 program during the 2022A semester (GS-2022A-Q-320). 
Each observation was performed in an ABBA nod sequential mode to efficiently remove sky background. 
Depending on the brightness of the target, one or two ABBA sequential observations were made. None of the observations lasted longer than 1200 seconds. 
The total exposure times for each star ranged from 360 to 2400 seconds. This was designed to achieve a signal-to-noise ratio (S/N) of approximately 100 per pixel. 
However, the actual S/N, estimated from the entire spectral region, varied from 50 to 120 depending on the star.


\subsection{Data reduction} \label{sec:sub:reduction}
The basic data reduction of the obtained spectra was conducted using the IGRINS Pipeline Package Version 2.2 \citep[PLP;][]{Lee2017}, which includes procedures such as flat-fielding correction, bad-pixel correction, subtraction of sky background, wavelength calibration, and aperture extraction. 
The PLP results in a one-dimensional spectrum from each ABBA mode observation. 
This spectrum comprises diffraction orders 98 to 122 in the H-band and orders 72 to 92 in the K-band.
We generated a continuous spectrum for each target by combining the effective wavelength ranges from each diffraction order \citep[see Table~2 of][]{Lim2022}. 
We also converted the wavelength scale from vacuum into air at this stage \citep{Morton2000}. 
Subsequently, continuum-normalized spectra were obtained through continuum fitting using the {\em continuum} task in the IRAF {\em onedspec} package.  

\begin{figure*}
\centering
   \includegraphics[width=1.0\textwidth]{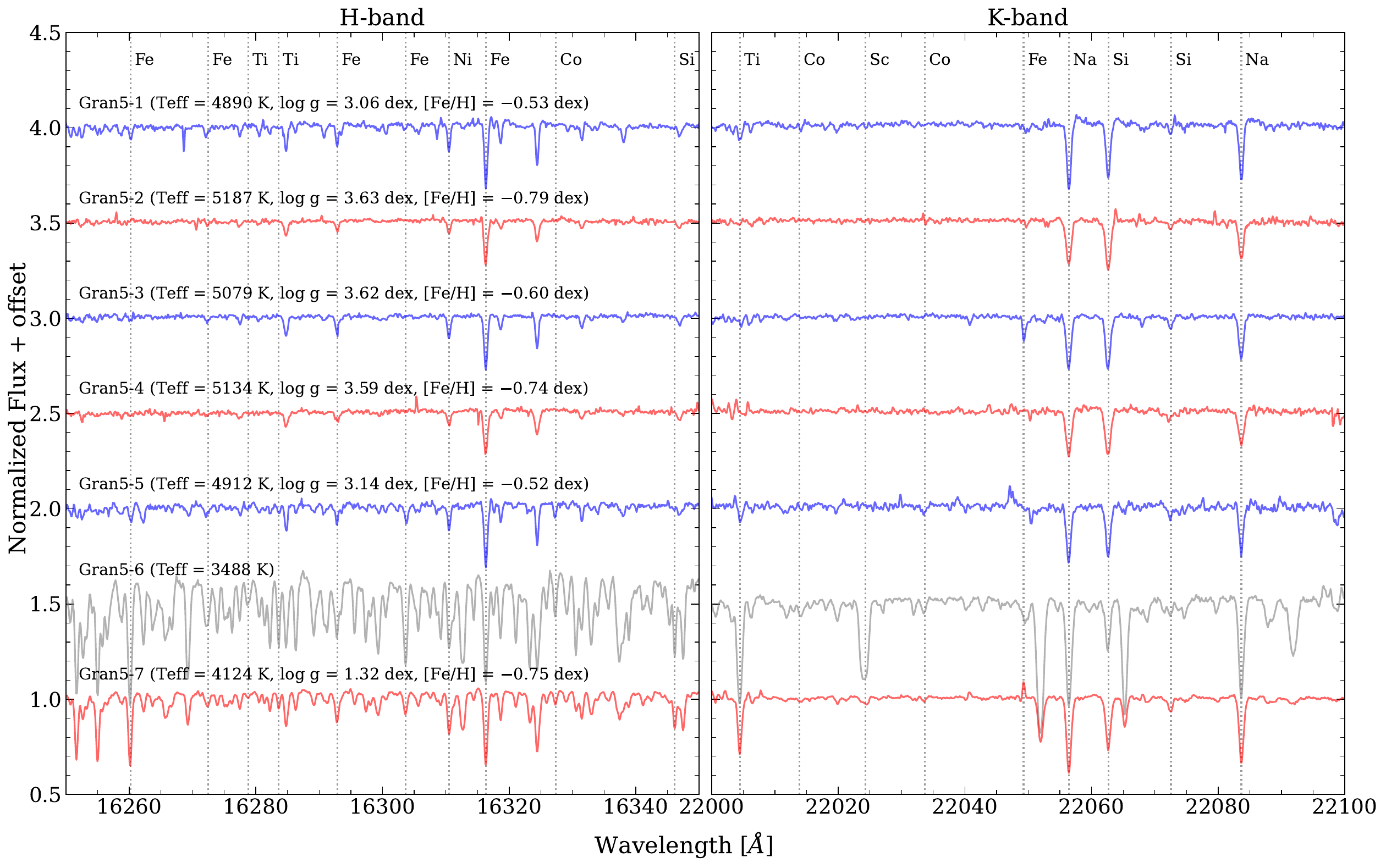}
     \caption{Continuum-normalized spectra for seven stars in the H- and K-band regions. 
     The chemical elements associated with each absorption line are labeled at the top of the figure.
     Gran~5-6 shows significantly different feature from the other stars.
     }
     \label{fig:spectra}
\end{figure*}

For each spectrum, we measured the heliocentric RV (RV$_{helio}$) using the {\em fxcor} and {\em rvcorrect} tasks in the IRAF {\em RV} package.
The synthetic spectrum from the Pollux database \citep{Palacios2010} was employed as a reference for the cross-correlation function.
Since there is a large gap between the H- and K-band spectral regions, we measured the RVs for each band separately and corrected them to the rest frame.
The RV$_{helio}$ and measurement errors listed in Table~\ref{tab:param} are derived from the arithmetic mean of the H- and K-band spectra and the propagation of each error.
For three stars (Gran~5-2, Gran~5-5, and Gran~5-7), which were observed with two ABBA sequences, the two spectra were combined after the RV correction.
The relatively large RV errors for these stars are due to the squared sum of four error values from the H- and K-band spectra of the two observations.
Figure~\ref{fig:spectra} shows a final spectrum for each star in the H- and K-band region, respectively. 

\begin{table*}
\caption{Heliocentric RV and atmospheric parameters}
\label{tab:param} 
\centering                                    
\begin{tabular}{cccccc}  
\hline\hline 
\multirow{2}{*}{ID} & RV$_{helio}$  & T$_{\rm eff}$ & $\log{g}$ & $v_{t}$       & [Fe/H]$_{model}$  \\ 
                    & [km~s$^{-1}$] & [K]           & [dex]     & [km~s$^{-1}$] & [dex]             \\
\hline
Gran~5-1     & $-$60.40 $\pm$ 0.51     & 4890 $\pm$ 73 & 3.06 $\pm$ 0.23 & 1.14 $\pm$ 0.07 & $-$0.53 $\pm$ 0.02 \\
Gran~5-2     & $-$61.69 $\pm$ 1.00     & 5187 $\pm$ 88 & 3.63 $\pm$ 0.19 & 1.04 $\pm$ 0.06 & $-$0.79 $\pm$ 0.01 \\
Gran~5-3     & $-$55.90 $\pm$ 0.52     & 5079 $\pm$ 74 & 3.62 $\pm$ 0.20 & 0.98 $\pm$ 0.07 & $-$0.60 $\pm$ 0.02 \\
Gran~5-4     & $-$60.74 $\pm$ 0.60     & 5134 $\pm$ 79 & 3.59 $\pm$ 0.20 & 1.03 $\pm$ 0.07 & $-$0.74 $\pm$ 0.02 \\
Gran~5-5     & $-$60.97 $\pm$ 0.93     & 4912 $\pm$ 74 & 3.14 $\pm$ 0.24 & 1.11 $\pm$ 0.08 & $-$0.52 $\pm$ 0.02 \\
Gran~5-6     & $-$9.55 $\pm$ 0.28      & 3488 $\pm$ 79 &            --   &              -- &                 -- \\
Gran~5-7     & $-$59.22 $\pm$ 1.00     & 4124 $\pm$ 84 & 1.32 $\pm$ 0.16 & 1.60 $\pm$ 0.04 & $-$0.75 $\pm$ 0.02 \\
\hline
\end{tabular}
\end{table*}

\section{Cluster membership and spectroscopic analysis} \label{sec:member_spec} 
\subsection{Cluster membership} \label{sec:sub:member}
As mentioned in Sect.~\ref{sec:intro}, the first step in studying a GC candidate is the spectroscopic confirmation and identification of the member stars. 
In Table~\ref{tab:param}, the RV$_{helio}$ value of Gran~5-6 is significantly different from those of the other stars, while the values for six stars are distributed within 5~km~s$^{-1}$.
This discrepancy indicates that Gran~5-6 is not a member star of Gran~5, despite the careful target selection based on MUSE observations (see Sect.~\ref{sec:sub:obs}). 
Although \citet{Gran2024} estimated an RV of $-$54.28~km~s$^{-1}$ for this star, which is similar to that of the other stars, it appears to be an incorrect estimate. 
The significantly different spectral features of Gran~5-6, as shown in Fig.~\ref{fig:spectra}, also support the claim that this star does not belong to the same group as the remaining six stars.
Therefore, we excluded this star from our analysis on the basis that Gran~5-6 is a field star.

The mean RV$_{helio}$ from six stars, excluding Gran~5-6, is estimated to be $-$59.82~km~s$^{-1}$ with a standard deviation of 2.08~km~s$^{-1}$. 
The RV of each star does not exceed 2$\sigma$ from the mean value. 
Because these stars were selected from a similar proper motion region \citep[see Fig. A1 of][]{Gran2022}, this narrow velocity dispersion strengthens the evidence that Gran~5 is a populated stellar cluster. 
\citet{Gran2024} estimated a mean RV of $-$59.19$\pm$4.93~km~s$^{-1}$ from 42 MUSE sample stars, which closely agrees with our measurement, although case-by-case discrepancies such as with Gran~5-6 may occur. 
The follow-up analysis was conducted on the basis that our six sample stars comprise a stellar cluster.


\subsection{Atmospheric parameters} \label{sec:sub:atm}
Prior to measuring chemical abundances, we estimated the atmospheric parameters, effective temperature (${\rm T_{eff}}$), surface gravity ($\log{g}$), microturbulence ($v_{t}$), and [Fe/H], for each star to generate a model atmosphere. 

We first measured ${\rm T_{eff}}$, the parameter that affects chemical abundance measurements most.
Although ${\rm T_{eff}}$ can be measured using the color-temperature relation, this method is unsuitable for stars toward the bulge because the distances are uncertain and the extinction is significant. 
We employed the line-depth-ratio (LDR) method, which compares the strength of high- and low-excitation lines, following previous NIR spectroscopy studies \citep{Fukue2015, BocekTopcu2020, Lim2022}. 
We individually measured ${\rm T_{eff}}$ from 35 line pairs available in our NIR spectral ranges, 9 pairs from \citet{Fukue2015} and 26 pairs from \citet{Afsar2023}. 
The final value of ${\rm T_{eff}}$ was estimated from the mean of these measurements, excluding some line pairs with low S/N or those outside a 3$\sigma$ distribution.

Our ${\rm T_{eff}}$ estimates obtained from the LDR method are typically 500~K higher than those reported by \citet{Gran2024}, who used the AMBRE grid of synthetic spectra \citep{deLaverny2012} on MUSE data.
These discrepancies critically influence the determination of other atmospheric parameters and chemical abundances. 
To verify the reliability of our ${\rm T_{eff}}$ estimates, we derived the photometric ${\rm T_{eff}}$ from the ($J - K_{s}$) color using the relation from \citet{GonzalezHernandez2009}, with reddening vectors from {\em Bayestar 19} \citep{Green2019} assuming a distance of 4.5~kpc \citep{Gran2022}.
Despite significant uncertainties from reddening, these photometric ${\rm T_{eff}}$ values agree better with our estimates. They show an average difference of 165~K from our LDR estimates and are higher by approximately 545~K than those of \citet{Gran2024}.
The ${\rm T_{eff}}$ values estimated using the equivalent width of the CO band \citep{Park2018} are also closer to our LDR method estimates, with variations ranging from 95~K to 260~K for three stars (4985~K for Gran~5-1, 4650~K for Gran~5-5, and 4241~K for Gran~5-7).
It is noteworthy that the CO band is not obvious in other three stars, which exhibit ${\rm T_{eff}}$ values higher than 5000~K according to our estimates.
The better consistency of our ${\rm T_{eff}}$ values with other estimates using different methods than those of \citet{Gran2024} supports the assumption that the ${\rm T_{eff}}$ obtained from the LDR method is reliable \citep[see also][]{Lim2022}.

The $\log{g}$, $v_{t}$, and [Fe/H] for the model atmosphere were simultaneously estimated at fixed ${\rm T_{eff}}$. 
We derived $\log{g}$ from an isochrone fitting on the Hertzsprung–Russell diagram, following the concept of \citet{Rich2017}, and we estimated $v_{t}$ using the empirical relation provided by \citet{Mashonkina2017}. 
It was challenging to determine $\log{g}$ for our sample stars. 
The canonical photometric method is limited due to uncertain distance information of bulge stars, and the spectroscopic method cannot be applied due to the absence of \ions{Fe}{ii} lines in the NIR spectral range. 
In our previous study, although we used the equivalent width of the CO-overtone band \citep{Lim2022}, this band was not observed in most of our samples due to their high temperatures.
Therefore, we fit Yonsei-Yale (Y$^{2}$) isochrones \citep{Kim2002} to our samples because our six stars are RGB stars of the same GC with the same age.  
We initially applied 10~Gyr $\alpha$-enhanced isochrones with an initial guess of $-$0.7~dex for [Fe/H]. 
The $\log{g}$ for each star was derived from the corresponding value to the ${\rm T_{eff}}$ of isochrones in the Hertzsprung–Russell diagram. 
After estimating $v_{t}$ with these parameters, we subsequently measured the [Fe/H] abundance ratio for each star through a spectral synthesis of a number of absorption lines. 
Finally, we iteratively measured the $\log{g}$, $v_{t}$, and [Fe/H] parameters for each star using newly interpolated isochrones with the newly derived [Fe/H] values until the output matched the input.

In order to examine the effect of the age of isochrones on the determination of $\log{g}$, we measured $\log{g}$ and [Fe/H] using isochrones of different ages, 3~Gyr, 7~Gyr, and 13~Gyr. 
The difference in $\log{g}$ between the maximum and minimum values was estimated to be up to 0.4~dex, with a typical standard deviation of 0.15~dex, leading to differences in [Fe/H] of only 0.01~dex to 0.03~dex. 
Thus, we confirmed that the effect of age is insignificant on the $\log{g}$ and chemical abundance measurement in the case of RGB stars in a GC, as reported by \citet{Rich2017}.
The minimal variation in [Fe/H] due to the age of isochrones ensures that any variations in the ages of our sample stars have a negligible impact on our chemical abundance measurements.

The uncertainties in the determination of atmospheric parameters were assessed sequentially. 
Because our ${\rm T_{eff}}$ estimates were derived from a large number of LDR pairs, the uncertainty of ${\rm T_{eff}}$ was first computed as the error of the mean from each LDR estimate.
We added an uncertainty of $\sim$70~K for each star. This reflects the uncertainty level of the LDR method suggested by \citet{Afsar2023}.
Similarly, the uncertainty of [Fe/H] was obtained from the error of the mean of [Fe/H] measured from each absorption line. 
We computed the uncertainty on $\log{g}$ through a Monte Carlo sampling (\(N=100,000\)), employing newly interpolated isochrones that account for the errors in both ${\rm T_{eff}}$ and [Fe/H].
We included the error in the $\log{g}$ determination due to the age uncertainty of employing isochrones, which was obtained from $\log{g}$ estimates using the four different ages mentioned above, during this Monte Carlo simulation. 
Finally, the uncertainty of $v_{t}$ was estimated using the same Monte Carlo sampling, incorporating all other uncertainties.

Table~\ref{tab:param} provides the final values of ${\rm T_{eff}}$, $\log{g}$, $v_{t}$, and [Fe/H] for the atmospheric model of each star, along with their corresponding uncertainties. 
In the case of Gran~5-6, which was excluded from cluster membership (see Sect.~\ref{sec:sub:member}), we measured only ${\rm T_{eff}}$. 
The other parameters could not be derived because its age and metallicity were presumed to differ significantly from those of the other stars.
 

\subsection{Chemical abundance measurement} \label{sec:sub:abund}
\begin{figure*}
\centering
   \includegraphics[width=0.95\textwidth]{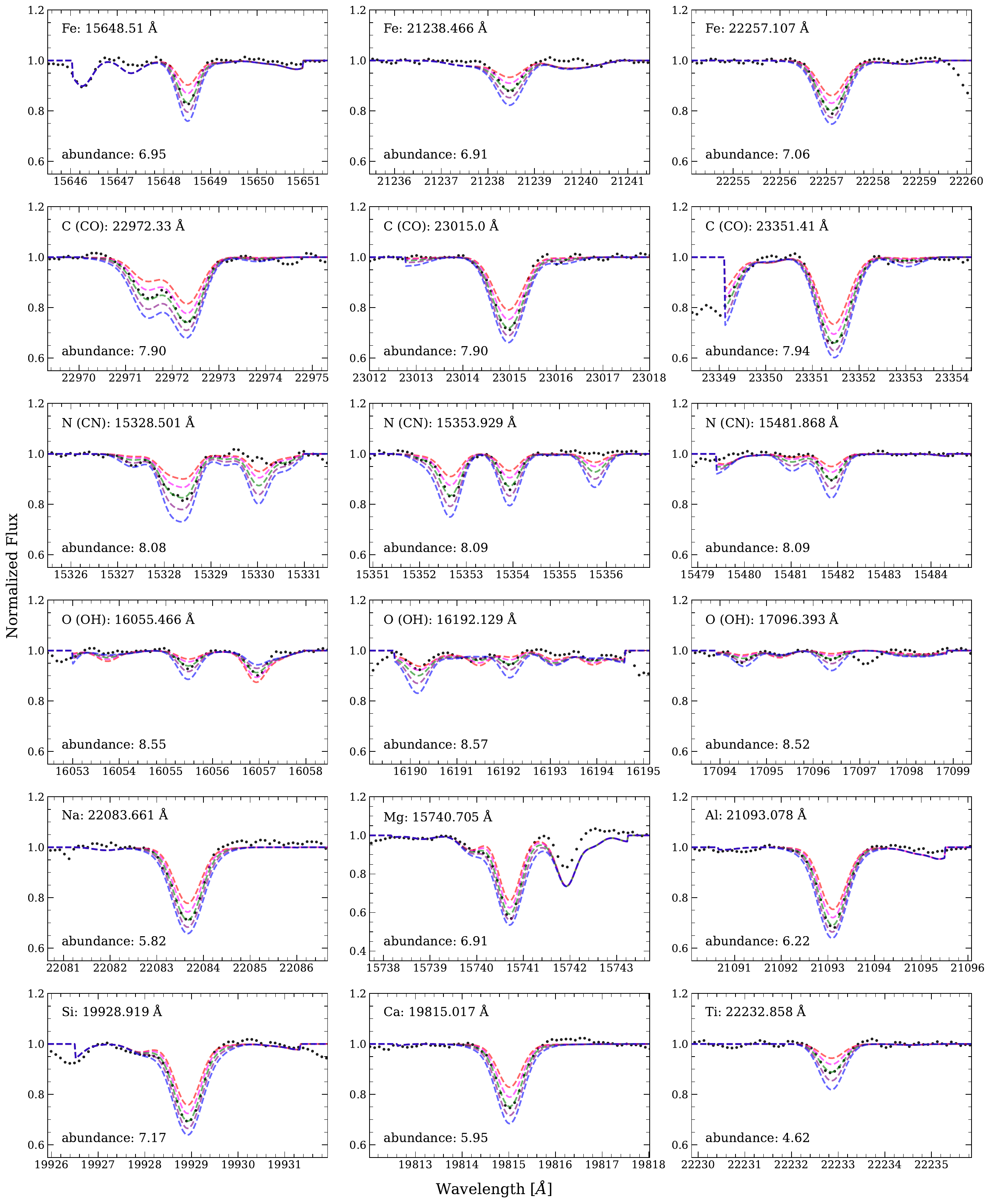}
     \caption{Observed spectrum of Gran~5-1 (black points) with synthetic spectra (dashed colored lines). The green line represents the best-fit model spectrum, and the other lines correspond to variations of $-0.4$, $-0.2$, $+0.2$, and $+0.4$ from this model.
     }
     \label{fig:synth}
\end{figure*}

To determine the chemical element abundances, we generated a model atmosphere for each star using a grid of plane-parallel MARCS models \citep{Gustafsson2008} for Gran~5-1 to Gran~5-5, while spherical MARCS models were employed for Gran~5-7 due to its low $\log{g}$. 
With these models, line-by-line chemical abundance measurements were conducted using the spectral synthesis method with the 2019NOV version of the local thermodynamic equilibrium (LTE) code MOOG \citep{Sneden1973}.
Figure~\ref{fig:synth} provides examples of the spectral synthesis process. 
The observed spectra were compared with synthetic spectra for each absorption feature to determine the best-fit abundance.

We initially measured the abundances of C, N, and O elements from the molecular features of the CO, CN, and OH bands available in our spectral range, with more than ten features for each band. 
Since these elements are interconnected, we iteratively measured each abundance in the order of O (OH), C (CO), and N (CN), updating the model atmosphere at each step with the newly estimated abundances of the other elements. 
After three iterations, we obtained fixed values for all three elements. 
However, the chemical abundances of C, N, and O could not be derived for Gran~5-2 and Gran~5-4 because the molecular features in these stars are weak. We attribute this to their high temperatures.

Subsequently, we measured the chemical abundances of 16 elements, Fe, Na, Mg, Al, Si, P, S, K, Ca, Sc, Ti, Cr, Co, Ni, Ce, and Yb, from absorption lines using a line list obtained from \citet{Lim2022}. 
The final values of chemical abundances were calculated as the mean of abundances estimated from individual lines for the same element. 
The chemical abundance ratios listed in Table~~\ref{tab:abund} were derived adopting the solar scale from \citet{Asplund2009}, and the measurement errors were estimated as $\sigma/\sqrt{N}$, where $\sigma$ is the standard deviation, and $N$ is the number of lines for each element.

\begin{table*}
\caption{Chemical abundance ratios and their measurement errors}
\label{tab:abund} 
\centering                                    
\begin{tabular}{cccccccc}  
\hline\hline 
\multirow{2}{*}{[X/H]} & \multirow{2}{*}{Gran~5-1} & \multirow{2}{*}{Gran~5-2} & \multirow{2}{*}{Gran~5-3} & \multirow{2}{*}{Gran~5-4} & \multirow{2}{*}{Gran~5-5} & \multirow{2}{*}{Gran~5-7} & Systematic \\
& & & & & & & Uncertainty \\
\hline
Fe & $-$0.53 $\pm$ 0.02 & $-$0.79 $\pm$ 0.01 & $-$0.60 $\pm$ 0.02 & $-$0.74 $\pm$ 0.02 & $-$0.52 $\pm$ 0.02 & $-$0.75 $\pm$ 0.02 & 0.05 \\
C  & $-$0.46 $\pm$ 0.02 & ... & $-$0.65 $\pm$ 0.01 & ... &    0.17 $\pm$ 0.04 & $-$0.43 $\pm$ 0.03 & 0.09 \\
N  &    0.23 $\pm$ 0.01 & ... &    0.00 $\pm$ 0.01 & ... & $-$0.61 $\pm$ 0.01 & $-$0.78 $\pm$ 0.02 & 0.06 \\
O  & $-$0.12 $\pm$ 0.02 & ... & $-$0.33 $\pm$ 0.02 & ... & $-$0.21 $\pm$ 0.02 & $-$0.21 $\pm$ 0.01 & 0.10 \\
Na & $-$0.42 $\pm$ 0.00 & $-$0.58 $\pm$ 0.01 & $-$0.47 $\pm$ 0.03 & $-$0.56 $\pm$ 0.02 & $-$0.51 $\pm$ 0.04 & $-$0.69 $\pm$ 0.01 & 0.08 \\
Mg & $-$0.57 $\pm$ 0.04 & $-$0.80 $\pm$ 0.03 & $-$0.61 $\pm$ 0.04 & $-$0.75 $\pm$ 0.03 & $-$0.42 $\pm$ 0.07 & $-$0.55 $\pm$ 0.0 & 0.08 \\
Al & $-$0.22 $\pm$ 0.06 & $-$0.37 $\pm$ 0.06 & $-$0.22 $\pm$ 0.06 & $-$0.30 $\pm$ 0.06 & $-$0.19 $\pm$ 0.07 & $-$0.37 $\pm$ 0.09 & 0.07 \\
Si & $-$0.37 $\pm$ 0.04 & $-$0.35 $\pm$ 0.03 & $-$0.26 $\pm$ 0.03 & $-$0.34 $\pm$ 0.04 & $-$0.30 $\pm$ 0.04 & $-$0.54 $\pm$ 0.05 & 0.03\\
P  &  0.26 $\pm$ ... &  0.26 $\pm$ ... &  0.25 $\pm$ ... & $-$0.24 $\pm$ ... &  0.40 $\pm$ ... & $-$0.55 $\pm$ ... & 0.07 \\
S  & $-$0.22 $\pm$ 0.02 &  0.06 $\pm$ 0.03 &  0.04 $\pm$ 0.03 &  0.03 $\pm$ 0.03 & $-$0.10 $\pm$ 0.04 & $-$0.21 $\pm$ 0.02 & 0.08 \\
K  & $-$0.58 $\pm$ 0.09 & $-$0.53 $\pm$ 0.01 & $-$0.10 $\pm$ 0.04 & $-$0.64 $\pm$ 0.05 & $-$0.43 $\pm$ 0.04 & $-$0.88 $\pm$ 0.05 & 0.05 \\
Ca & $-$0.40 $\pm$ 0.02 & $-$0.60 $\pm$ 0.02 & $-$0.35 $\pm$ 0.03 & $-$0.54 $\pm$ 0.02 & $-$0.39 $\pm$ 0.07 & $-$0.36 $\pm$ 0.02 & 0.06 \\
Sc &  0.11 $\pm$ ... & $-$0.56 $\pm$ ... &  0.41 $\pm$ ... &   ... & $-$0.38 $\pm$ ... & $-$0.69 $\pm$ 0.05 & 0.10 \\
Ti & $-$0.28 $\pm$ 0.03 & $-$0.63 $\pm$ 0.08 & $-$0.45 $\pm$ 0.04 & $-$0.61 $\pm$ 0.07 & $-$0.26 $\pm$ 0.02 & $-$0.47 $\pm$ 0.11 & 0.10 \\
Cr & $-$0.77 $\pm$ ... & $-$0.80 $\pm$ 0.04 & $-$0.78 $\pm$ 0.04 & $-$0.92 $\pm$ 0.06 & $-$0.85 $\pm$ ... & $-$0.85 $\pm$ ... & 0.07 \\
Co & $-$0.91 $\pm$ ... & $-$1.70 $\pm$ ... & $-$1.14 $\pm$ ... & $-$0.62 $\pm$ ... &   ... & $-$1.21 $\pm$ ... & 0.12 \\
Ni & $-$0.53 $\pm$ 0.05 & $-$0.71 $\pm$ 0.05 & $-$0.60 $\pm$ 0.04 & $-$0.55 $\pm$ 0.06 & $-$0.56 $\pm$ 0.11 & $-$0.76 $\pm$ 0.13 & 0.04 \\
Ce & $-$0.30 $\pm$ 0.05 & $-$0.30 $\pm$ 0.02 & $-$0.49 $\pm$ 0.04 & $-$0.22 $\pm$ 0.02 &  0.92 $\pm$ 0.03 & $-$1.07 $\pm$ 0.15 & 0.13 \\
Yb & $-$0.02 $\pm$ ... &   ... & $-$0.01 $\pm$ ... & $-$0.13 $\pm$ ... &  0.52 $\pm$ ... &   ... & 0.09\\
\hline
\end{tabular}
\tablefoot{The statistical measurement error could not be obtained when the abundance ratios were measured from a single absorption line.
Column 8 provides the typical systematic uncertainties for each element due to the uncertainties in the determination of the atmospheric parameters.}
\end{table*}

In addition, we evaluated the systematic uncertainties affecting chemical abundance measurements, which can arise from uncertainties in the determination of atmospheric parameters (see Table~\ref{tab:param}). 
As a case study, we remeasured the abundances of Gran~5-2 ($\sigma_{\rm T_{eff}}$ = $\pm$88~K, $\sigma_{\log{g}}$ = $\pm$0.19~dex, $\sigma_{v_{t}}$ = $\pm$0.06~km~s$^{-1}$, and $\sigma_{\rm [Fe/H]}$ = $\pm$0.01~dex) and Gran~5-3 ($\sigma_{\rm T_{eff}}$ = $\pm$74~K, $\sigma_{\log{g}}$ = $\pm$0.20~dex, $\sigma_{v_{t}}$ = $\pm$0.07~km~s$^{-1}$, and $\sigma_{\rm [Fe/H]}$ = $\pm$0.02~dex) using eight alternative atmospheric models.
Each model was adjusted both upward and downward in response to the uncertainties in ${\rm T_{eff}}$, $\log{g}$, $v_{t}$, and [Fe/H]. 
We then measured the differences in the abundance from the original values listed in Table~\ref{tab:abund} for each element.
Of the four atmospheric parameters, the uncertainty on ${\rm T_{eff}}$ affects the abundance measurements most. 
For instance, the systematic error in the [Fe/H] ratio due to the ${\rm T_{eff}}$ uncertainty is $\pm$0.05~dex, while the errors due to the $\log{g}$, $v_{t}$, and [Fe/H] uncertainties are negligible, each is smaller than $\pm$0.01~dex.
However, the abundance ratios of P, S, Ce, and Yb are more significantly impacted by the uncertainties on $\log{g}$. 
For elements measured from molecular bands, the systematic errors in the C and N abundance ratios due to the $\log{g}$ uncertainty are comparable to those caused by the uncertainties in ${\rm T_{eff}}$, whereas the O abundance ratio is primarily influenced by the uncertainties on ${\rm T_{eff}}$.
The upper limit of the total systematic uncertainty was determined by the squared sum of the uncertainties for the four parameters.
The total systematic uncertainties for each element, calculated as the average of two measurements for Gran~5-2 and Gran~5-3, are presented in Table~\ref{tab:abund}.


\section{Chemical properties of the stars in Gran~5} \label{sec:result_chem} 
\subsection{Two stellar populations with different metallicities} \label{sec:sub:Fe}
\begin{figure}
\centering
   \includegraphics[width=0.5\textwidth]{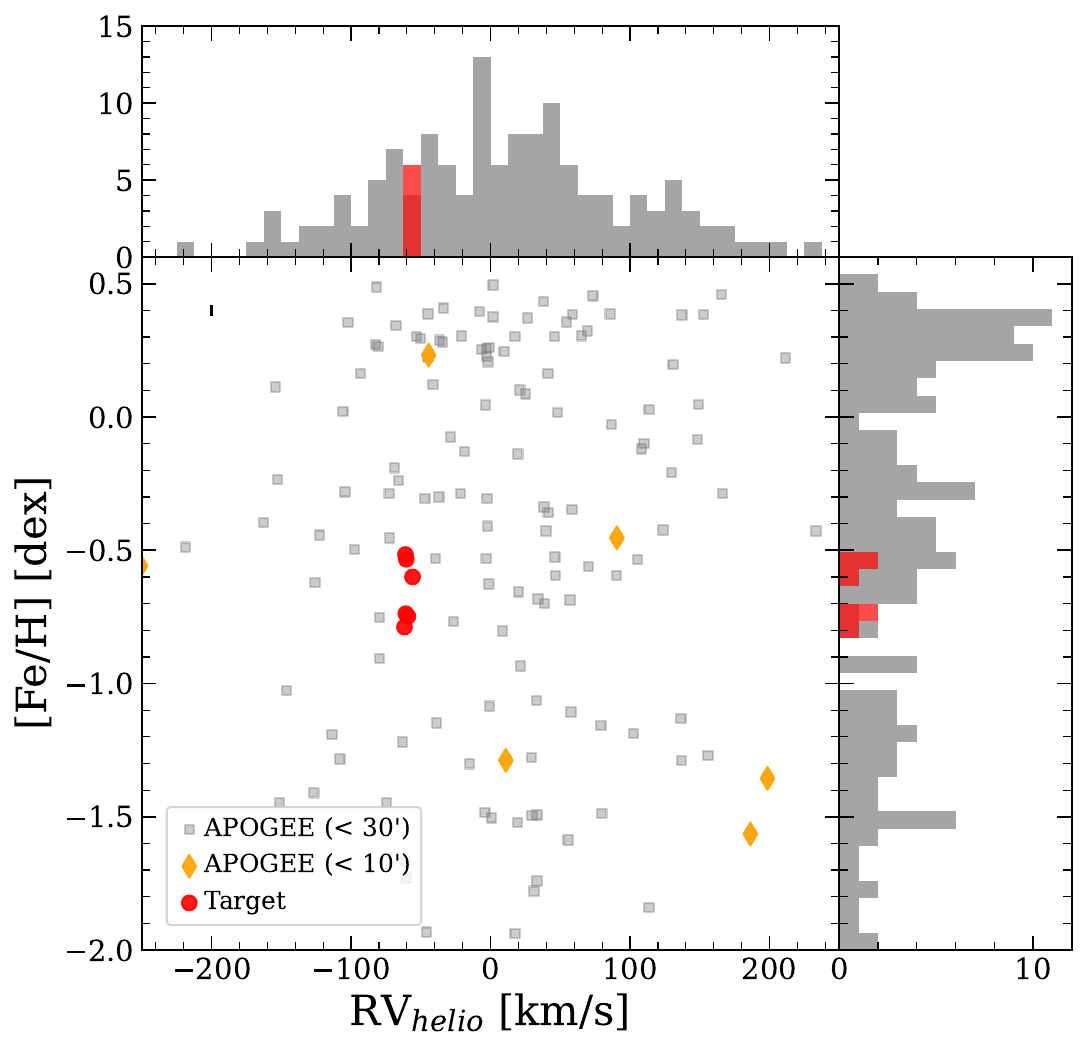}
     \caption{Comparison of our target stars (red circles) to the nearby field stars from APOGEE (gray squares) on the [Fe/H] - RV$_{helio}$ plane.
     The upper and right panels show histograms of RV$_{helio}$ and [Fe/H], respectively.
     While our target stars fall into the same bin in the histogram of RV$_{helio}$, they can be divided into two metallicity groups in the histogram of [Fe/H].
     The mean error is plotted in the upper left corner, although it is comparable to the marker size.  
     The yellow diamonds represent APOGEE sample stars located within 10$\arcmin$ from the center of Gran~5.
     However, none of these stars appears to be a member of Gran~5 based on their position in this plane.
     }
     \label{fig:FeH_RV}
\end{figure}

In the previous section, we measured the chemical abundance ratios for six stars that were confirmed as member stars of Gran~5 based on their close RV$_{helio}$ similarity, in addition to their location and proper motion.
The metallicity of these stars ranges from $-0.79$ to $-0.52$~dex in [Fe/H]. 
As shown in Fig.~\ref{fig:FeH_RV}, our six sample stars are distinct from nearby field stars on the [Fe/H] - RV$_{helio}$ plane, supporting the fact that Gran~5 is a stellar cluster. 
The mean [Fe/H] value is estimated to be $-0.65$~dex, with a standard deviation of 0.11~dex. 
Although this metallicity places Gran~5 on the metal-rich side of all GCs in the Milky Way \citep{Harris2010}, a significant number of bulge GCs are found within the [Fe/H] range of $-0.7$~dex to $-0.5$~dex \citep[see][]{Bica2024}. 
Thus, the metallicity of Gran~5 derived from our sample stars suggests the possibility that it is a GC that belongs to the Galactic bulge.

Interestingly, our six target stars can be divided into two subpopulations with different metallicties, as shown in Fig.~\ref{fig:FeH_RV}. 
The metal-poor group consists of three stars of Gran~5-2, Gran~5-4, and Gran~5-7, and the other three stars of Gran~5-1, Gran~5-3, and Gran~5-5 belong to the metal-rich group. 
The mean [Fe/H] value is $-0.76$~dex for the metal-poor group and $-0.55$~dex for the metal-rich group, with standard deviations of 0.02~dex and 0.04~dex, respectively.
The difference in [Fe/H] between the two groups is estimated to be 0.21~dex.
Considering the small typical measurement error in [Fe/H] ($\sim$0.02~dex) and the standard error of the mean for each group ($\sim$0.02~dex), this difference is statistically significant.
The substantial Kullback-Leibler (KL) divergence of 47.5 between the Gaussian mixture models (GMMs) fit to the metal-poor and metal-rich populations further supports the assumption that these two groups originate from distinct distributions.
Furthermore, these two populations do not show any tendency with RV$_{helio}$ or atmospheric parameters.
Therefore, despite the limited sample size, we conclude that Gran~5 likely consists of two populations with different metallicities, similar to the cases of M22 \citep{Marino2011} and NGC~6273 \citep{Johnson2015}, for example. 
However, it is necessary to examine whether stars in Gran~5 are clearly divided into two metallicity groups or if they show a wide range of variation in [Fe/H] with a larger sample size. 

On the other hand, our measurements of the [Fe/H] ratios are higher than those reported by \citet{Gran2024}, who indicated a mean [Fe/H] of $-1.02$~dex for Gran~5. 
Individually, the differences range from 0.16~dex (Gran~5-2) to 0.63~dex (Gran~5-5), with our estimates being higher. 
This discrepancy might be caused by the different wavelength coverages and spectral resolutions used in the two observations with MUSE and IGRINS, respectively, as well as variations in spectroscopic analysis methods. 
In particular, significant differences in ${\rm T_{eff}}$ may influence the [Fe/H] measurements (see Sect.~\ref{sec:sub:atm}). 
The stars with larger differences in ${\rm T_{eff}}$ show larger discrepancies in [Fe/H], as exemplified by Gran~5-5 ($\Delta$[Fe/H] = 0.63~dex with $\Delta{\rm T_{eff}} = 582$~K) and Gran~5-2 ($\Delta$[Fe/H] = 0.16~dex with $\Delta{\rm T_{eff}} = 345$~K).
We expect that our [Fe/H] ratios are more accurate because our determination of ${\rm T_{eff}}$ appears to be more reliable, as discussed in Sect.~\ref{sec:sub:atm}, and our chemical abundances are measured at a higher spectral resolution of $R \sim 45,000$.


\subsection{Other chemical abundance patterns} \label{sec:sub:Na_O}
\begin{figure}
\centering
   \includegraphics[width=0.37\textwidth]{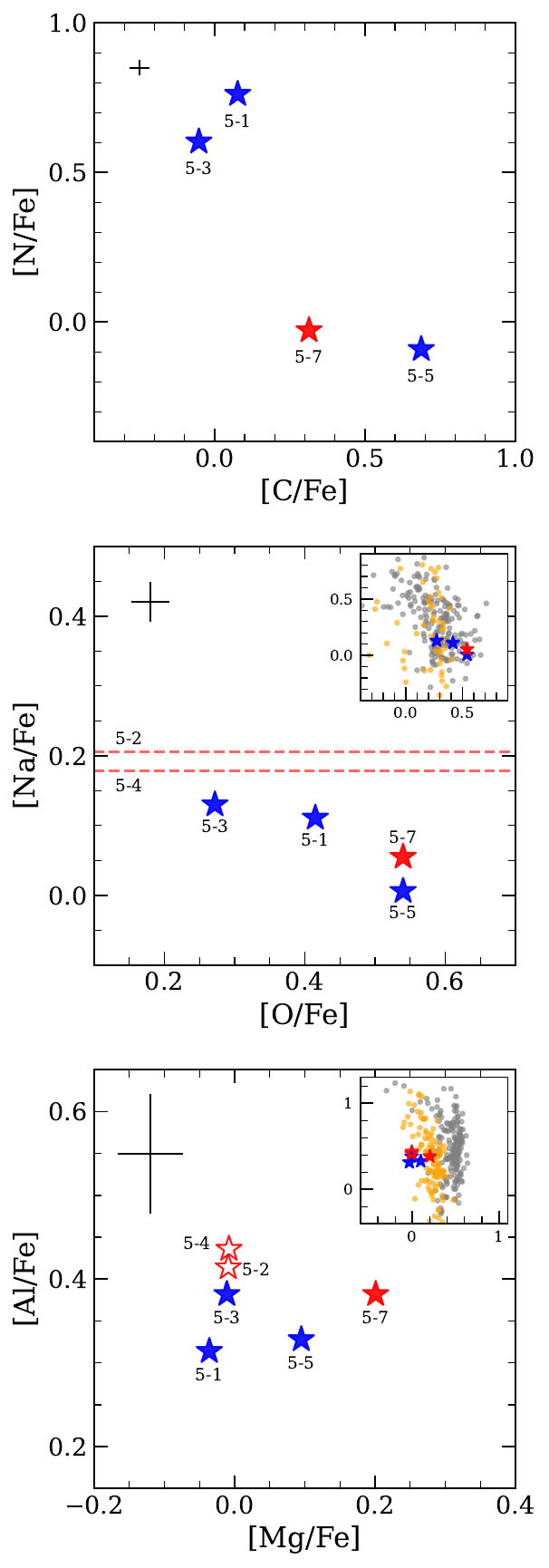}
     \caption{Distributions of the stars in Gran~5 in the [N/Fe] vs. [C/Fe] (upper), [Na/Fe] vs. [O/Fe] (middle), and [Al/Fe] vs. [Mg/Fe] (lower) plots. 
     The blue and red symbols indicate metal-rich and metal-poor populations, respectively, as divided in Sect.~\ref{sec:sub:Fe}. 
     Gran~5-2 and Gran~5-4 are represented as open symbols only in the lower panel, and their [Na/Fe] abundance ratios are indicated by dashed red lines in the middle panel because we lack C, N, and O abundance data. 
     The mean measurement error for each element ($\pm1\sigma$) is plotted in the upper left corner of each panel. 
     The inset panels in the middle and lower panels provide zoomed-out views for comparison with stars in other GCs. 
     The gray circles represent data from 17 GCs derived from optical spectroscopy \citep{Carretta2009}, and the yellow circles correspond to stars in NGC~6304 and NGC~6273 obtained from the APOGEE catalog by \citet{Schiavon2024}.
     }
     \label{fig:Na_O}
\end{figure}

A significant chemical signature of GCs is the chemical anomalies between multiple stellar populations, such as the Na-O and Mg-Al anticorrelations \citep[see, e.g.,][]{Carretta2009, Bastian2018}. 
These anomalies are observed in most Milky Way GCs, regardless of whether they host subpopulations with different metallicities. 
We compare the chemical patterns of these elements in Fig.~\ref{fig:Na_O}, which illustrates C-N, Na-O, and Mg-Al anticorrelations.

In these figures, no clear distinction is observed between the metal-poor and metal-rich populations. 
This is consistent with other GCs, which show metallicity variations but no distinct separation between metal-poor and metal-rich populations in light elements, while still exhibiting the Na-O anticorrelation within each group \citep[see, e.g.,][]{Marino2011, Marino2015}. 
In this respect, it appears that our sample stars also show some chemical anomalies in these elements, regardless of their metallicity.
Gran~5-1 and Gran~5-3 are more enhanced in N, but depleted in C compared to Gran~5-5 and Gran~5-7\footnote{Although evolutionary mixing can deplete C and enhance N with increasing luminosity during the RGB phase \citep[e.g.,][]{Placco2014}, this effect cannot account for our results, as the stars in our sample have comparable luminosities (see Fig.~\ref{fig:cmd}).},
as shown in the upper panel of Fig.~\ref{fig:Na_O}. 
In addition, these two stars are slightly enhanced in Na and depleted in O and Mg, while no enhancement in Al is observed. 
Conversely, Gran~5-5 and Gran~5-7 are enriched in O and Mg, but depleted in Na compared to Gran~5-1 and Gran~5-3. 
Gran~5-2 and Gran~5-4, whose chemical abundances of C, N, and O were not derived (see Sect.~\ref{sec:sub:abund}), show enhancements in Na and Al and depletion in Mg compared to the other stars. 
Based on these characteristics, we infer that Gran~5-5 and Gran~5-7 represent a primordial stellar population, whereas Gran~5-1, Gran~5-2, Gran~5-3, and Gran~5-4 likely belong to a chemically enriched stellar population as is typically observed in GCs.

However, it is challenging to assert that our sample stars of Gran~5 provide definitive evidence for the presence of multiple stellar populations as observed in other GCs. 
As shown in the insets of Fig.~\ref{fig:Na_O}, our samples are far more narrowly distributed than the stars from \citet{Carretta2009}, which demonstrate Na-O and Mg-Al anticorrelations in 17 GCs. 
Thus, although the chemical patterns observed in our data are qualitatively similar to those of general GCs with multiple stellar populations, their distributions are quantitatively insufficient to represent these features.
One important consideration in the comparison between our data and those of \citet{Carretta2009} is the systematic discrepancy in the chemical abundance measurements that may arise from our spectroscopic analysis in the NIR region.
Most previous high-resolution spectroscopic studies of GC stars have been conducted in the optical wavelength region.
Unlike optical stellar spectroscopy, the accurate properties of absorption lines, including non-LTE effects, are less studied in the NIR region. 
For instance, as shown in the inset in the lower panel of Fig.~\ref{fig:Na_O}, our [Mg/Fe] abundance ratios are systematically lower than those reported by \citet{Carretta2009}.
This underestimate of [Mg/Fe] measured in the NIR region has been reported in previous studies \citep{Nandakumar2023, Lim2024}. 
Moreover, APOGEE NIR data for inner Galactic GCs (NGC~6304 and NGC~6273) similarly indicate that the Mg abundance ratios derived from the NIR range are generally lower than those obtained from the optical band (see the yellow circles in  Fig.~\ref{fig:Na_O}).
Therefore, although the distinct Na-O and Mg-Al anticorrelations observed in other GCs are not evident in Gran~5, it remains challenging at this stage to determine whether the observed chemical patterns in Gran~5 are due to multiple populations within the GC or are a systematic bias in our NIR spectroscopic analysis.
Nevertheless, since the chemical abundances of the six sample stars are self-consistently derived from the same observation and method, their relative chemical patterns are significant for understanding the stellar populations of Gran~5.


\section{Chemodynamical properties of Gran~5} \label{sec:orbit}
It is a significant task as part of understanding the assembly history of the Milky Way to classify Galactic GCs into either in situ or accreted objects and to identify the associated accretion events for the latter \citep{Callingham2022, Belokurov2024}.
\citet{Gran2022} suggested that Gran~5 is a bulge GC that is associated with the $Gaia$-Enceladus structure based on its orbital parameters. 
However, subsequent revisions by \citet{Gran2024} of the RV measurements and the revision of the [Fe/H] ratios in this study necessitate a re-evaluation of its origin using updated chemodynamical information.

\begin{figure}
\centering
   \includegraphics[width=0.46\textwidth]{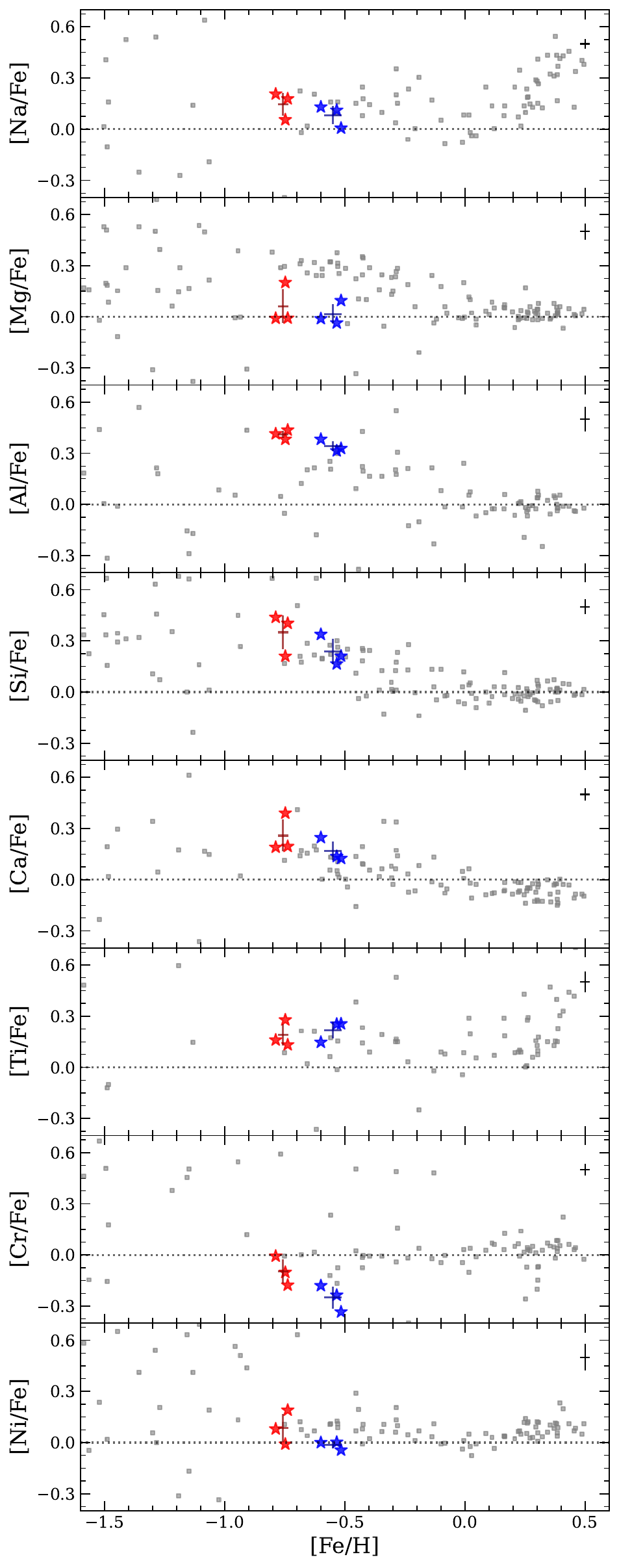}
     \caption{Chemical abundance ratios of the target stars, together with nearby APOGEE sample stars that are located within 30$\arcmin$ from the center of Gran~5 (gray squares). 
     Metal-rich and metal-poor stars are represented by blue and red symbols, respectively,
     and the mean abundance ratios and standard deviations for each group are marked as blue and red crosses.
     The black cross in the upper right corner of each panel indicates the mean measurement error ($\pm1\sigma$) from the six sample stars, and the dotted horizontal lines represent the value of 0.0~dex for each abundance ratio.
     }
     \label{fig:abund}
\end{figure}

\begin{table}
\caption{Mean abundance ratios and standard deviations for two metallicity groups}
\label{tab:mean} 
\centering                                    
\begin{tabular}{cccccc}  
\hline\hline 
\multirow{2}{*}{[X/Fe]}& \multicolumn{2}{c}{metal-poor}    & \multicolumn{2}{c}{metal-rich} & \multirow{2}{*}{$\Delta$}  \\ 
\cline{2-3} \cline{4-5} 
                       & $\mu$         & $\sigma$          & $\mu$         & $\sigma$       &           \\
\hline
Na                     & $0.15$        & 0.07              & $0.08$        & 0.05           & $0.07$    \\
Mg                     & $0.06$        & 0.10              & $0.02$        & 0.06           & $0.04$    \\
Al                     & $0.41$        & 0.02              & $0.34$        & 0.03           & $0.07$    \\
Si                     & $0.35$        & 0.10              & $0.24$        & 0.07           & $0.11$    \\
Ca                     & $0.26$        & 0.09              & $0.17$        & 0.06           & $0.09$    \\
Ti                     & $0.19$        & 0.06              & $0.22$        & 0.05           & $-0.03$   \\
Cr                     & $-0.10$       & 0.07              & $-0.25$       & 0.06           & $0.15$    \\
Ni                     & $0.09$        & 0.08              & $-0.01$       & 0.02           & $0.10$    \\
\hline
\end{tabular}
\tablefoot{The difference between the two groups listed in Column 6 is calculated as the value for the metal-poor group subtracted by that for the metal-rich group.}
\end{table}

To examine the chemical properties of Gran~5, we compared our data to nearby samples from APOGEE DR17 \citep{Abdurro'uf2022}. 
Within a 30$\arcmin$ radius from the center of Gran~5, 189 stars were identified, with [Fe/H] ratios ranging from $-2.0$ to $+0.5$~dex. 
Despite the proximity of 10 stars within 10$\arcmin$ of the center, none of these stars shows RV$_{helio}$ and [Fe/H] values that are comparable to our samples (see Fig.~\ref{fig:FeH_RV}). 
Figure~\ref{fig:abund} presents a comparison of the chemical abundance ratios of eight elements Na, Mg, Al, Si, Ca, Ti, Cr, and Ni, excluding elements with relatively large uncertainties or without APOGEE measurements. 
In this figure, the metal-poor and metal-rich groups are clearly distinguished. The mean abundance ratios and standard deviations for each group are listed in Table~\ref{tab:mean}.

The overall chemical properties of all sample stars agree with those of the APOGEE sample despite the methodological differences in data analysis. 
We note that stellar parameters and chemical abundance ratios from the APOGEE survey are obtained by comparing observed spectra to a library of synthetic spectra in the H-band wavelength region with a zeropoint shift. 
However, our estimates of [Mg/Fe] and [Cr/Fe] abundance ratios appear to be underestimated compared to the APOGEE sample, as discussed in Sect.~\ref{sec:sub:Na_O}. 
The application of a non-LTE effect correction for Cr \citep{Bergemann2010} resulted in an increase by 0.16~dex in its ratio. 
For Mg, the mean abundance increased by 0.09~dex when we included certain absorption lines in the K-band region, which were excluded because their values were relatively high compared to other measurements.

Consequently, our sample stars in Gran~5 show no peculiar chemical signature compared to nearby field stars. 
No evidence of $\alpha$-element depletion was found for Si, Ca, and Ti, except for Mg, which is characteristic of accreted objects \citep[e.g.,][]{Nissen2010, Lim2021a}. 
However, the absence of $\alpha$-element depletion could be due to the relatively higher metallicity of Gran~5, as this depletion is not evident in the metal-rich stars.  
The most distinct finding of this study from \citet{Gran2022} is the higher metallicity of Gran~5, revised from $-1.56$~dex to $-0.65$~dex (see Sect.~\ref{sec:sub:Fe}). This questions its classification as a GC associated with the $Gaia$-Enceladus structure. 
\citet{Koch-Hansen2021} compared the [$\alpha$/Fe] abundance ratios of $Gaia$-Enceladus-related GCs as a function of [Fe/H] (see their Fig.~10), and \citet{Callingham2022} classified 23 GCs in the $Gaia$-Enceladus group (see their Fig.~9).
Compared to these studies, Gran~5 is far more metal-rich than other GCs that are dynamically associated with $Gaia$-Enceladus ([Fe/H] $\lesssim$ $-1.0$~dex), while \citet{Helmi2018} reported $Gaia$-Enceladus stars approximately spanning from $-2.5$ to $-0.5$~dex in [Fe/H], peaking at $-1.6$~dex. 
Thus, the derived chemical properties more likely support the idea that Gran~5 is an in situ GC and do not support the accreted origin associated with $Gaia$-Enceladus or other structures \citep[see also][]{Belokurov2024}.

\begin{figure}
\centering
   \includegraphics[width=0.48\textwidth]{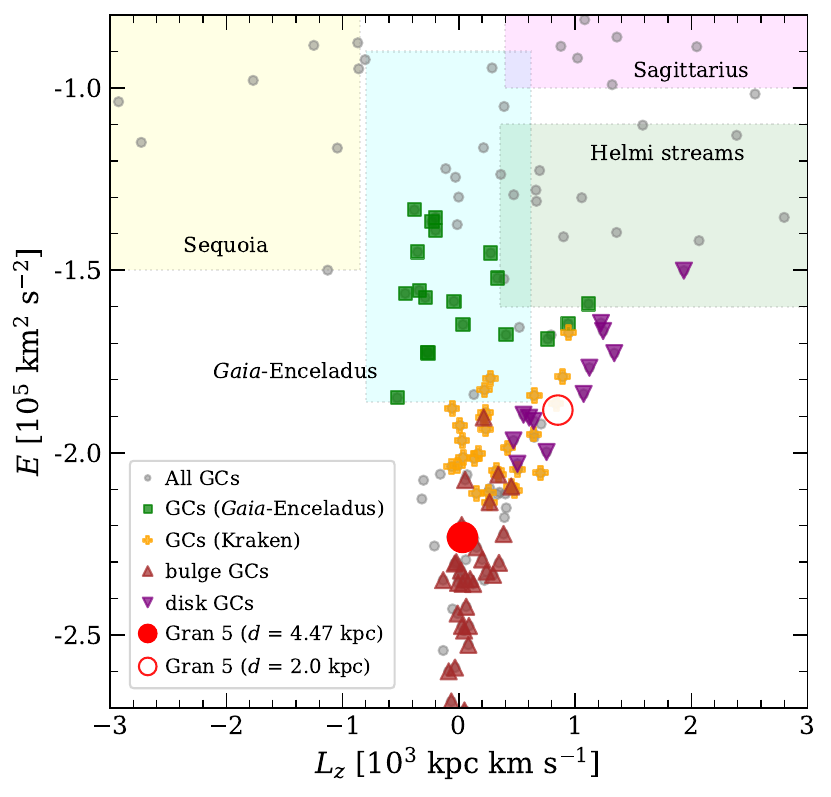}
     \caption{$E$ -- $L_{Z}$ diagram for Milky Way GCs, including Gran~5. 
     The classifications of the GCs follow the grouping by \citet{Callingham2022}, represented with different colors and symbols as listed in the lower left corner.
     The colored background boxes denote the dynamical domains of the Sequoia, $Gaia$-Enceladus, Helmi-streams, and Sagittarius, adapted from \citet{Massari2019}. 
     We illustrate the position of Gran~5 in this diagram for two different distances: 4.47~kpc and 2.0~kpc. 
     }
     \label{fig:orbit}
\end{figure}

\begin{figure*}
\centering
   \includegraphics[width=0.8\textwidth]{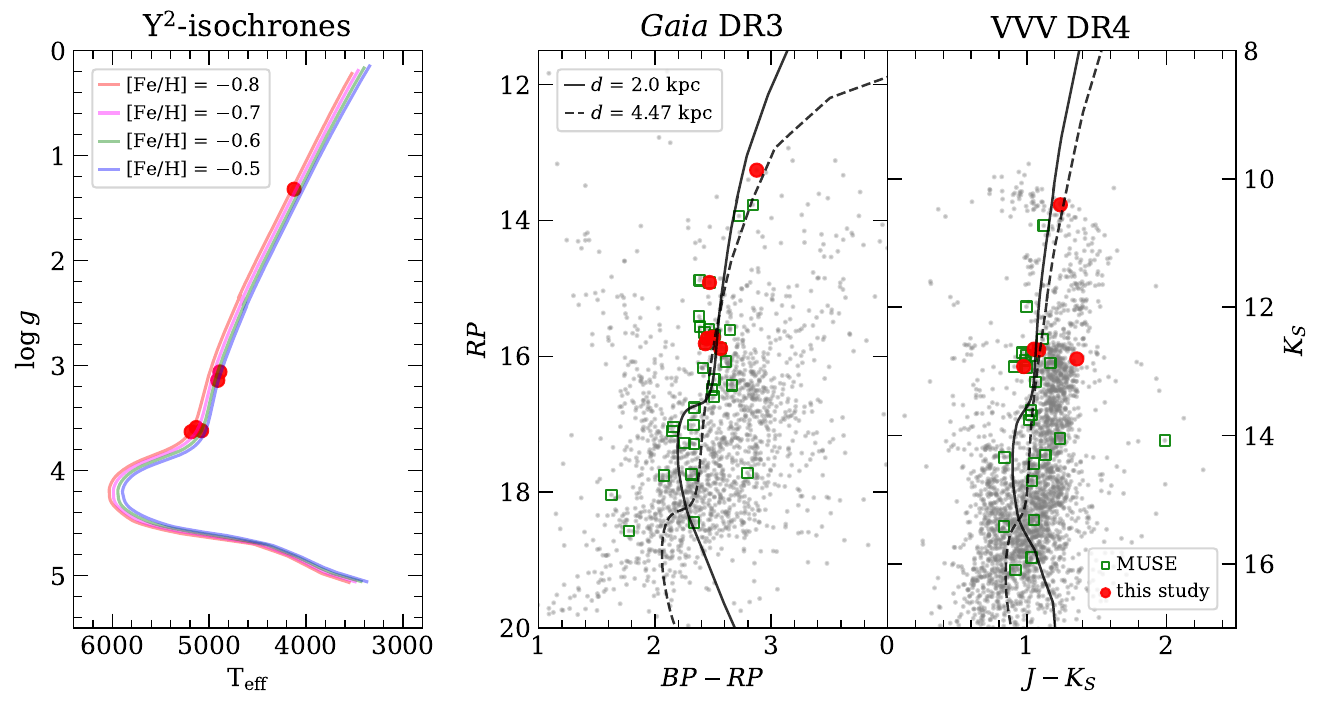}
     \caption{Application of isochrones in the Kiel diagram (left panel), and two CMDs from $Gaia$ DR3 and VVV DR4 data (right panel). 
     The left panel shows the determined ${\rm T_{eff}}$ and $\log{g}$ parameters from Sect.~\ref{sec:sub:atm} for our sample stars, overlaid with 10~Gyr Y$^{2}$ isochrones for four different metallicities.
     The CMDs in the right panel are identical to those in Fig.~\ref{fig:cmd}, but we plot isochrones of $-0.7$~dex in [Fe/H].
     We applied a distance of 2.0~kpc and $E(B-V)$ of 1.1~mag for the solid line and a distance of 4.47~kpc and $E(B-V)$ of 1.0~mag for the dashed line.
     Although both lines closely follow the RGB sequence, the locations of our sample stars on the RGB differ. They appear as faint and bright RGB for the 2.0~kpc and 4.47~kpc lines, respectively.
     The location of the sample stars on the isochrone at 2.0~kpc is more comparable to their positions in the Kiel diagram. 
     }
     \label{fig:iso}
\end{figure*}

A more distinct comparison was conducted in the orbital parameter plane. 
We calculated the orbital energy ($E$) and angular momentum ($L_{Z}$) for Gran~5 using the {\em galpy} package \citep{Bovy2015} with the \texttt{McMillan2017} potential \citep{McMillan2017}. 
These calculations were based on the mean values of R.A., decl., and proper motion ($\mu_{\alpha}$ and $\mu_{\delta}$) for six target stars obtained from $Gaia$ DR3 and the mean RV derived in this study ($-59.82$~km~s$^{-1}$). 
For the distance to Gran~5, which is the most uncertain parameter, a distance of 4.47~kpc was initially applied, referencing \citet{Gran2022}. 
Figure~\ref{fig:orbit} displays the $E$ and $L_{Z}$ parameters for Milky Way GCs, including Gran~5. 
As shown in the figure, Gran~5 (filled red circle) is positioned differently from other GCs that are dynamically associated with the $Gaia$-Enceladus structure. 
While a more detailed study of the orbital trajectory of Gran~5 is required, the current data suggest that it likely belongs to the bulge GC group in the $E$ -- $L_{Z}$ diagram.

In addition, we re-estimated $E$ and $L_{Z}$ for Gran~5, assuming a closer distance of 2.0~kpc (represented by the open red circle in Fig.~\ref{fig:orbit}). 
This revised assumption was based on the higher ${\rm T_{eff}}$ of our sample stars compared to those reported by \citet{Gran2024} (see Sect.~\ref{sec:sub:atm}). 
The increased ${\rm T_{eff}}$ suggests that these stars are situated in the fainter region of the RGB and not in the brighter region, as illustrated in the Kiel diagram (the left panel of Fig.~\ref{fig:iso}). 
The fitting of a 10~Gyr Y$^{2}$ isochrone with an [Fe/H] of $-0.7$~dex to the CMDs supports the hypothesis that a distance of 2~kpc more accurately reflects the location of our sample stars on the isochrone corresponding to the Kiel diagram\footnote{However, since our analysis is based on only six stars, a more detailed fitting for a larger verified sample of member stars is required in both the Kiel diagram and CMDs to determine the precise distance of Gran~5.}. 
When we adopted a distance of 2~kpc for Gran~5, its orbital parameters became more comparable to those of the disk GCs in Fig.~\ref{fig:orbit}.

Consequently, regardless of whether a distance of 4.47~kpc or 2.0~kpc is adopted, no evidence is found to support a dynamical association of Gran~5 with the $Gaia$-Enceladus structure.
At a distance of 4.47~kpc, Gran~5 shows characteristics typical of in situ bulge GCs, whereas at a distance of 2.0~kpc, it more closely resembles disk GCs. 
This result agrees with the classification by \citet{Belokurov2024}, who identified Gran~5 as an in situ  GC.
If Gran~5 is an accreted object, its orbital parameters agree more closely with the Kraken structure \citep{Kruijssen2019} than with $Gaia$-Enceladus, as shown in the $E$ -- $L_{Z}$ diagram of Fig.~\ref{fig:orbit}. 
The Kraken structure is identified as the low-energy group by \citet{Massari2019} and as the inner Galaxy group (or Heracles) by \citet{Horta2021}.
In this study, we refer to it as Kraken, following the classification by \citet{Callingham2022}.
However, the overall chemodynamical properties of Gran~5 suggest that it is more likely an in situ GC that belongs to the Galactic bulge or disk.


\section{Discussion} \label{sec:discussion} 
Based on the high-resolution NIR spectroscopy of seven stars located in the field of Gran~5, we identified six stars as cluster members, while one star was excluded due to its significantly different RV. 
The confirmation that the RVs of the six stars are similar ($\overline{\rm RV_{\it helio}}$ = $-$59.82$\pm$2.08~km~s$^{-1}$) and their low metallicities ($\overline{\rm [Fe/H]}$ = $-$0.65$\pm$0.11~dex) substantiates the claim that Gran~5 is indeed a GC.
However, these stars can be divided into two stellar populations with different metallicities, with mean [Fe/H] values of $-0.76$~dex and $-0.55$~dex for each group. 
In addition, the chemodynamical properties of Gran~5 agree more closely with those of in situ GCs that belong to the Galactic bulge or disk than with accreted GCs that are associated with the $Gaia$-Enceladus or Kraken structures. 

Although there are several GCs with metallicity variations in the Milky Way, Gran~5 is distinct from the others. 
Multiple stellar populations with different metallicities have been reported in GCs such as M2 \citep{Yong2014}, M22 \citep{Marino2011}, NGC~5286 \citep{Marino2015}, and NGC~6273 \citep{Johnson2015}. 
Since the metallicity variation in a GC is thought to be due to enrichment by type II supernovae, a higher initial mass is required for these GCs to retain supernova ejecta. 
These types of GCs are typically more massive, with present-day masses greater than $4 \times 10^5$ M$_{\odot}$ and initial masses expected to exceed $10^6$ M$_{\odot}$ (\citealt{Baumgardt2018}\footnote{Referenced from the 4th version of the catalog: \url{https://people.smp.uq.edu.au/HolgerBaumgardt/globular/}}).
In contrast, the present-day mass of Gran~5, estimated at only $2.29 \times 10^4$ M$_{\odot}$ from the same catalog, is significantly lower than that of the previously mentioned GCs.
This significant discrepancy suggests that the observed metallicity variation of Gran~5 may have arisen from different processes, or that it lost a substantial amount of mass during its evolution. 
Both scenarios seem plausible given the distinct chemodynamics of Gran~5. 
The mean [Fe/H] of Gran~5 is relatively higher at $-0.65$~dex than for other GCs with metallicity variations \citep[$<$ $-1.5$~dex;][]{Harris2010}. 
It is important to note that metallicity variations are also found in metal-rich bulge GCs such as Terzan~5 \citep{Ferraro2009, Origlia2011} and Liller~1 \citep{Crociati2023}. 
Although the chemical evolution of multiple populations in these GCs is expected to be more complex with considerable differences in metallicity and age between populations, it is possible that Gran~5 lies between the metal-rich bulge GCs and other metal-poor GCs, both of which exhibit metallicity variations. 
Alternatively, given that Gran~5 is located within the Galactic plane at $l$ = 4.459$\degree$ and $b$ = 1.838$\degree$, it may have been significantly affected by the Galactic structure and other GCs, leading to the loss of a substantial portion of its initial mass \citep{Ishchenko2023a, Ishchenko2023b}.
This hypothesis suggests that Gran~5 and similar GCs could have significantly contributed to the formation of the Galactic bulge.

On the other hand, we cannot completely rule out the possibility that the differences in metallicity between the two populations could be attributed to systematic errors in the determination of atmospheric parameters and chemical abundances from our spectroscopy.
Some studies have shown that metallicity variations in certain GCs can be mitigated by applying $\log{g}$ values derived from photometric methods instead of the canonical spectroscopic method \citep{Mucciarelli2015, Lardo2016}. 
In our study, the derivation of atmospheric parameters from standard methods is restricted due to the absence of \ions{Fe}{ii} lines in the NIR region and the large uncertainties in magnitude and distance caused by the severe extinction of stars toward the Galactic bulge \citep[see also][]{Lim2022}. 
Therefore, a further high-resolution spectroscopic analysis for a larger sample of stars from Gran~5 through various techniques is  essential to clearly confirm these metallicity variations.

The importance of high-resolution NIR spectroscopy becomes more significant as the number of new objects discovered in the Galactic plane increases. 
Furthermore, investigating the chemical properties in the NIR band of distant stars in nearby galaxies will become increasingly crucial with the advent of extremely large telescopes.
However, stellar high-resolution NIR spectroscopy and the necessary validation against previous optical studies remain insufficient. 
In this context, the recent commissioning of the NIR spectrometer IGRINS-2 on the Gemini-North telescope \citep{Diaz2022, Lee2022} can serve as a bridge to future NIR stellar spectroscopic studies. 
We also plan to perform high-resolution NIR spectroscopy of both previously known and newly reported GCs in the Milky Way.

\begin{acknowledgement}
We are grateful to the referee for a number of helpful suggestions.
D.L., Y.W.L., and C.C. acknowledge support from the National Research Foundation of Korea (NRF) to the Center for Galaxy Evolution Research (2022R1A6A1A03053472 and 2022R1A2C3002992). 
S.H.C. acknowledges support from the NRF grant funded by the Korea government (MSIT) (NRF-2021R1C1C2003511) and the Korea Astronomy and Space Science Institute under R\&D program (Project No. 2024-1-831-00) supervised by the Ministry of Science and ICT.
S.H. acknowledges support from the LAMP Program of the NRF grant funded by the Ministry of Education (No. RS-2023-00301976).
D.L. thanks Sree Oh for the ongoing support.
This work was supported by K-GMT Science Program (PID: GS-2022A-Q-320, GS-2023A-Q-126, and GS-2023A-Q-216) of Korea Astronomy and Space Science Institute (KASI).
This work used the Immersion Grating Infrared Spectrometer (IGRINS) that was developed under a collaboration between the University of Texas at Austin and the KASI with the financial support of the Mt. Cuba Astronomical Foundation, of the US National Science Foundation under grants AST-1229522 and AST-1702267, of the McDonald Observatory of the University of Texas at Austin, of the Korean GMT Project of KASI, and Gemini Observatory.    
\end{acknowledgement}

\bibliographystyle{aa} 
\bibliography{export-bibtex}

\end{document}